\documentclass[12pt,english]{article}
\usepackage{geometry}
\usepackage{amsmath}
\usepackage{amssymb}

\makeatletter
\newcommand{\lyxaddress}[1]{
\par {\raggedright #1
\vspace{1.4em}
\noindent\par}
}

\usepackage{babel}
\makeatother
\begin{document}

\title{Electromagnetic Duality, Quaternion and Supersymmetric Gauge Theories
of Dyons}

\author{H. Dehnen~ and ~O. P. S. Negi%
\thanks{Permanent Address:- Department of Physics, Kumaun University, S. S.
J. Campus ALMORA- 263601 (UA) India.%
}\\
}

\maketitle

\lyxaddress{\begin{center}Universität Konstanz\\
Fachbereich Physik\\
Postfach M 677\\
D-78457 Konstanz,Germany\\
Email:-Heinz.Dehnen@uni-konstanz.de\\
ops\_negi@yahoo.co.in\par\end{center}}

\begin{abstract}
Starting with the generalized potentials, currents, field tensors
and electromagnetic vector fields of dyons as the complex complex
quantities with real and imaginary counter parts as electric and magnetic
constituents, we have established the electromagnetic duality for
various fields and equations of motion associated with dyons in consistent
way. It has been shown that the manifestly covariant forms of generalized
field equations and equation of motion of dyons are invariant under
duality transformations. Quaternionic formulation for generalized
fields of dyons are developed and corresponding field equations are
derived in compact and simpler manner. Supersymmetric gauge theories
are accordingly reviewed to discuss the behaviour of dualities associated
with BPS mass formula of dyons in terms of supersymmetric charges.
Consequently, the higher dimensional supersymmetric gauge theories
for N=2 and N=4 supersymmetries are analysed over the fields of complex
and quaternions respectively.
\end{abstract}

\section{Introduction}

~~~The asymmetry between electricity and magnetism became very
clear at the end of 19th century with the formulation of Maxwell's
equations. Magnetic monopoles were advocated \cite{key-1} to symmetrize
these equations in a manifest way that the mere existence of an isolated
magnetic charge implies the quantization of electric charge and accordingly
the considerable literature \cite{key-2,key-3,key-4,key-5,key-6,key-7}
has come in force. The fresh interests in this subject have been enhanced
by 't Hooft \cite{key-8} and Polyakov \cite{key-9} with the idea
that the classical solutions having the properties of magnetic monopoles
may be found in Yang - Mills gauge theories. Now it has become clear
that monopoles are better understood in grand unified theories and
supersymmetric gauge theories. Julia and Zee \cite{key-10} extended
the ' t Hooft-Polyakov theory of monopoles and constructed the theory
of non Abelian dyons (particles carrying simultaneously electric and
magnetic charges). The quantum mechanical excitation of fundamental
monopoles include \cite{key-11} dyons which are automatically arisen
from the semi-classical quantization of global charge rotation degree
of freedom of monopoles. In view of the explanation of CP-violation
in terms of non-zero vacuum angle of world \cite{key-12}, the monopoles
are necessary dyons and Dirac quantization condition permits dyons
to have analogous electric charge. Renewed interests in the subject
of monopole has gathered enormous potential importance in connection
of quark confinement problem \cite{key-13} in quantum chromo dynamics,
possible magnetic condensation \cite{key-14} of vacuum leading to
absolute color confinement in QCD, its role as catalyst in proton-decay
\cite{key-15}, CP-violation \cite{key-12} , current grand unified
theories \cite{key-16} and supersymmetric gauge theories \cite{key-17,key-18,key-19,key-20}. 

There has been a revival in the formulation of natural laws within
the frame work of general quaternion algebra and basic physical equations.
Quaternions \cite{key-21} were very first example of hyper complex
numbers having the significant impacts on mathematics and  physics.
Moreover,quaternions are already used in the context of special relativity
\cite{key-22}, electrodynamics \cite{key-23,key-24}, Maxwell's equation
\cite{key-25}, quantum mechanics \cite{key-26,key-27}, quaternion
oscillator \cite{key-28}, gauge theories \cite{key-29,key-30}, supersymmetry
\cite{key-31} and  other branches of Physics \cite{key-32} and Mathematics
\cite{key-33}. On the other hand supersymmetry is described as the
symmetry of bosons and fermions \cite{key-34} . Gauge Hierarchy problem,
not only suggests that the SUSY exists but put an upper limit on the
masses of super partners \cite{key-35,key-36}. The exact SUSY implies
exact fermion-boson masses, which has not been observed so far. Hence
it is believed that supersymmetry is an approximate symmetry and it
must be broken \cite{key-37,key-38}. The quaternionic formulation
of generalized electromagnetic fields of dyons \cite{key-39} has
been developed \cite{key-40,key-41,key-42} in unique, simple, compact
and consistent manner. Postulation of Heavisidian monopole \cite{key-43}
immediately follows the structural symmetry between generalized gravito-Heavisidean
and generalized electromagnetic fields of dyons \cite{key-44}.

In order to understand the theoretical existence of monopoles (dyons)
and keeping in view their recent potential importance and the fact
that the formalism necessary to describe them has been clumsy and
is not manifestly covariant, in the present paper, we have revisited
the generalized fields of dyons, electromagnetic duality, quaternion
formulation, non -Abelian and supersymmetric gauge theories in a consistent
manner. Starting with the idea of two four potentials and taking the
generalized charge, current, field,potential and electromagnetic field
tensor as complex quantities with real and imaginary parts of them
as electric and magnetic constituents, we have discussed the manifestly
covariant and dual invariant field theory of generalized electromagnetic
fields of dyons. The electromagnetic duality between electric and
magnetic constituents of dyons have been also established in terms
of a duality matrix and Generalized Dirac Maxwell's (GDM) equations,
equation of motion ,energy and momentum densities associated with
dyons are shown to be invariant under duality transformations. It
has been analysed that the dual invariance is maintained only with
the introduction of magnetic charge and the present model has been
shown to be consistent for point like monopoles too and the Chirality
quantization parameter plays an important role in electromagnetic
duality to obtain the Dirac quantization condition. Quaternion formulation
of various field equations has also been discussed in terms its simple,
compact and consistent analyticity and it has been shown that the
quaternion field equations are not only covariant, it is invariant
under quaternion transformations of different kinds and represents
the self dual structure of dyonic fields. It has also been described
the model of quaternion fields of dyons is valid in classical,quantum,
gauge and supersymmetric theories of particles carrying simultaneously
electric and magnetic charges. Quaternion Dirac equation of dyons
is also established in terms of quaternion covariant derivative and
the gamma matrices associated therein are also described as quaternion
valued. It has been shown that the commutator of quaternion derivative
yields two gauge field strengths giving rise to the field equations
respectively associated with the presence electric and magnetic charges
and quaternion gauge theory extends theory of monopoles and dyons
to the non -Abelian gauge theories. It has also been discussed that
quaternion gauge theory can be dealt accordingly with different kinds
of gauge transformations and theory of non Abelian monopoles ('t Hooft
Polyakov model) and BPS model can be explained and reformulated in
terms of quaternion gauges. Constancy condition may be imposed accordingly
to relate electric and magnetic parameters of dyons in order to avoid
the introduction of another photons. The BPS mass formula and electromagnetic
duality are also explained in the frame work of quaternion gauge theory.
Accordingly supersymmetric gauge theories, Bogomolny bound , Mantonen-Olive
duality and $SL(2,\mathbb{Z})$ transformations of supersymmetric
gauge theories of dyons are revisited in the context of quaternion
representation. Consequently, the higher dimensional supersymmetric
gauge theories for N=2 and N=4 supersymmetries are also analysed respectively
over the fields of complex and quaternion hyper complex numbers. Finally,
it has been concluded that the quaternion formulation be adopted in
a better way to understand the explanation of the duality conjecture
and supersymmetric gauge theories as the candidate for the existence
of monopoles and dyons where the complex parameters be described as
the constituents of quaternion.

\section{Definitions of Quaternion}

The algebra $\mathbb{H}$ of quaternion is a four-dimensional algebra
over the field of real numbers $\mathbb{R}$ and a quaternion $\phi$
is expressed in terms of its four base elements as

\begin{equation}
\phi=\phi_{\mu}e_{\mu}=\phi_{0}+e_{1}\phi_{1}+e_{2}\phi_{2}+e_{3}\phi_{3}(\mu=0,1,2,3),\label{eq:1}\end{equation}
where $\phi_{0}$,$\phi_{1}$,$\phi_{2}$,$\phi_{3}$ are the real
quartets of a quaternion and $e_{0},e_{1},e_{2},e_{3}$ are called
quaternion units and satisfies the following relations,

\begin{eqnarray}
e_{0}^{2} & =e_{0}= & 1\,\,,e_{j}^{2}=-e_{0},\nonumber \\
e_{0}e_{i}=e_{i}e_{0} & = & e_{i}(i=1,2,3)\,\,,\nonumber \\
e_{i}e_{j} & = & -\delta_{ij}+\varepsilon_{ijk}e_{k}(\forall\, i,j,k=1,2,3)\,\,,\label{eq:2}\end{eqnarray}
where $\delta_{ij}$ is the delta symbol and $\varepsilon_{ijk}$
is the Levi Civita three index symbol having value $(\varepsilon_{ijk}=+1)$
for cyclic permutation , $(\varepsilon_{ijk}=-1)$ for anti cyclic
permutation and $(\varepsilon_{ijk}=0)$ for any two repeated indices.
Addition and multiplication are defined by the usual distribution
law $(e_{j}e_{k})e_{l}=e_{j}(e_{k}e_{l})$ along with the multiplication
rules given by equation (\ref{eq:2}). $\mathbb{H}$ is an associative
but non commutative algebra. If $\phi_{0},\phi_{1},\phi_{2},\phi_{3}$
are taken as complex quantities, the quaternion is said to be a bi-
quaternion. Alternatively, a quaternion is defined as a two dimensional
algebra over the field of complex numbers $\mathbb{C}$. We thus have
$\phi=\upsilon+e_{2}\omega(\upsilon,\omega\in\mathbb{C})$ and $\upsilon=\phi_{0}+e_{1}\phi_{1}$
, $\omega=\phi_{2}-e_{1}\phi_{3}$ with the basic multiplication law
changes to $\upsilon e_{2}=-e_{2}\bar{\upsilon}$.The quaternion conjugate
$\overline{\phi}$ is defined as 

\begin{equation}
\overline{\phi}=\phi_{\mu}\bar{e_{\mu}}=\phi_{0}-e_{1}\phi_{1}-e_{2}\phi_{2}-e_{3}\phi_{3}\,\,.\label{eq:3}\end{equation}
In practice $\phi$ is often represented as a $2\times2$ matrix $\phi=\phi_{0}-i\,\vec{\sigma}\cdot\vec{\phi}$
where $e_{0}=I,e_{j}=-i\,\sigma_{j}(j=1,2,3)$ and $\sigma_{j}$are
the usual Pauli spin matrices. Then $\overline{\phi}=\sigma_{2}\phi^{T}\sigma_{2}$
with $\phi^{T}$ the trans pose of $\phi$. The real part of the quaternion
$\phi_{0}$ is also defined as

\begin{eqnarray}
Re\,\phi & = & \frac{1}{2}(\overline{\phi}+\phi)\,\,,\label{eq:4}\end{eqnarray}
where $Re$ denotes the real part and if $Re\,\phi=0$ then we have
$\phi=-\overline{\phi}$ and imaginary $\phi$ is known as pure quaternion
written as

\begin{equation}
\phi=e_{1}\phi_{1}+e_{2}\phi_{2}+e_{3}\phi_{3}\,\,.\label{eq:5}\end{equation}
The norm of a quaternion is expressed as $N(\phi)=\phi\overline{\phi}=\overline{\phi}\phi=\sum_{j=0}^{3}\phi_{j}^{2}$which
is non negative i.e.

\begin{equation}
N(\phi)=\left|\phi\right|=\phi_{0}^{2}+\phi_{1}^{2}+\phi_{2}^{2}+\phi_{3}^{2}=Det.(\phi)\geq0.\label{eq:6}\end{equation}
Since there exists the norm of a quaternion, we have adivision i.e.
every $\phi$ has an inverse of a quaternion and is described as

\begin{equation}
\phi^{-1}=\frac{\overline{\phi}}{\left|\phi\right|}\,\,\,.\label{eq:7}\end{equation}
While the quaternion conjugation satisfies the following property

\begin{equation}
\overline{\phi_{1}\phi_{2}}=\overline{\phi_{2}}\,\overline{\phi_{1}}\,\,\,.\label{eq:8}\end{equation}
The norm of the quaternion (\ref{eq:6}) is positive definite and
enjoys the composition law

\begin{equation}
N(\phi_{1}\phi_{2})=N(\phi_{1})N(\phi_{2})\,\,\,.\label{eq:9}\end{equation}
Quaternion (\ref{eq:1}) is also written as $\phi=(\phi_{0},\vec{\phi})$
where $\vec{\phi}=e_{1}\phi_{1}+e_{2}\phi_{2}+e_{3}\phi_{3}$ is its
vector part and $\phi_{0}$ is its scalar part. So, the sum and product
of two quaternions are described as

\begin{eqnarray}
(\alpha_{0}\vec{,\,\alpha})+(\beta_{0}\vec{,\,\beta}) & = & (\alpha_{0}+\beta_{0},\,\vec{\alpha}+\vec{\beta})\,\,\,,\nonumber \\
(\alpha_{0}\vec{,\,\alpha})\cdot(\beta_{0}\vec{,\,\beta}) & = & (\alpha_{0}\beta_{0}-\overrightarrow{\alpha}\cdot\overrightarrow{\beta}\,,\alpha_{0}\overrightarrow{\beta}+\beta_{0}\overrightarrow{\alpha}+\overrightarrow{\alpha}\times\overrightarrow{\beta})\,\,.\label{eq:10}\end{eqnarray}
Quaternion elements are non-Abelian in nature and thus represent a
non commutative division ring.

\section{Fields Associated with Dyons}

Starting with the idea of Cabbing and Ferrari \cite{key-45} of two
four-potentials, a self-consistent, gauge covariant and Lorentz invariant
quantum field theory of dyons have been developed \cite{key-39,key-40,key-41,key-42}
on assuming the generalized charge, generalized current and generalized
four-potential of dyons as a complex quantity with their real and
imaginary parts as a electric and magnetic constituents i.e.

\begin{eqnarray}
q & = & e\,-i\, g\;\,\!\,\,\,\,\,\,\,(\, generalized\; charge)\,\,\,(i=\sqrt{-1}),\nonumber \\
J_{\mu} & = & J_{\mu}-i\, k_{\mu}\negthickspace\,\,\quad\,(\, generalized\; four-current\,)\,\,\,,\nonumber \\
V_{µ} & = & A_{µ}-i\, B_{µ}\,\qquad(\, generalized\; four-potential\,)\,\,,\label{eq:11}\end{eqnarray}
where $e$ is the electric charge, $g$ is the magnetic charge, $j_{µ}$
is the electric four-current, $k_{µ}$ is the magnetic four-current,
$A_{µ}$ is the electric four-potential and $B_{µ}$ is the magnetic
four-potential. introduction of second four potential gives rise to
the removal of arbitrary string variables \cite{key-1} . Singleton
\cite{key-46} described the necessity of second potential and called
it as pseudo vector (hidden) potential to formulate the symmetric
theory of classical electrodynamics in presence of electric and magnetic
charges while the Hamiltonian formulation of the Theory containing
electric and magnetic charges is also described by Barker-Granziani
\cite{key-47} . We recall now the symmetric Maxwell's equations derived
earlier \cite{key-1} as the Generalized Dirac Maxwell's (or GDM)
equations for dyons given as

\begin{eqnarray}
\overrightarrow{\nabla}\cdot\overrightarrow{E} & =\rho_{e}\,\,,\,\,\,\,\,\,\,\,\,\,\,\,\,\,\,\,\overrightarrow{\nabla}\times\overrightarrow{H}= & \overrightarrow{j}\,+\frac{\partial\overrightarrow{E}}{\partial t}\,\,,\nonumber \\
\overrightarrow{\nabla}\cdot\overrightarrow{H} & =\rho_{g}\,\,,\,\,\,\,\,\,\,\,\,\,\,\,\,\,\,\,\overrightarrow{\nabla}\times\overrightarrow{E}= & -\overrightarrow{k}\,-\frac{\partial\overrightarrow{H}}{\partial t}\,\,,\label{eq:12}\end{eqnarray}
where $\overrightarrow{E}$ is the electric field, $\overrightarrow{H}$
is the magnetic field, $\rho_{e}$ is the charge source density due
to electric charge , $\rho_{m}$ is the charge source density due
to magnetic charge (monopole), $\overrightarrow{j}\,$ is the current
source density due to electric charge and $\overrightarrow{k}\,$
is the current source density due to magnetic charge. We have used
the notations $\left\{ j^{^{µ}}\right\} =\left\{ \rho_{e},\overrightarrow{j}\,\right\} $,
$\left\{ k^{^{µ}}\right\} =\left\{ \rho_{m},\overrightarrow{k}\right\} $,
$\left\{ A^{^{µ}}\right\} =\left\{ \phi_{e},\overrightarrow{A}\right\} $
and $\left\{ B^{^{µ}}\right\} =\left\{ \phi_{g},\overrightarrow{B}\right\} $
along with the metric as $(+,-,-,-)$. We have also used the unit
value of coefficients along with natural units $c=\hbar=1$ through
out the text. The electric and magnetic fields of dyons satisfying
the GDM equations (\ref{eq:12}) are now expressed in terms of components
of two four potentials in a symmetrical manner i.e.

\begin{eqnarray}
\overrightarrow{E} & = & -\overrightarrow{\nabla}\phi_{e}-\frac{\partial\overrightarrow{A}}{\partial t}-\overrightarrow{\nabla}\times\overrightarrow{B}\,\,,\nonumber \\
\overrightarrow{H} & = & -\overrightarrow{\nabla}\phi_{g}-\frac{\partial\overrightarrow{B}}{\partial t}+\overrightarrow{\nabla}\times\overrightarrow{A}\,\,.\label{eq:13}\end{eqnarray}
The vector field $\overrightarrow{\psi}$ associated with generalized
electromagnetic fields of dyons is defined as

\begin{eqnarray}
\overrightarrow{\psi} & = & \overrightarrow{E}\,-\, i\:\overrightarrow{H}\label{eq:14}\end{eqnarray}
which reduces the four GDM equations (\ref{eq:12}) to following two
differential equations as

\begin{eqnarray}
\overrightarrow{\nabla}\cdot & \overrightarrow{\psi} & =\rho,\nonumber \\
\overrightarrow{\nabla}\times & \overrightarrow{\psi} & =-\, i\,\overrightarrow{J\,}-i\,\frac{\partial\overrightarrow{\psi}}{\partial t}\label{eq:15}\end{eqnarray}
where $\rho=\rho_{e}-i\,\rho_{g}$ is the generalized charge and $\overrightarrow{J\,}=\overrightarrow{j\,}-i\,\overrightarrow{k\,}$
is the generalized current source densities of dyons. These are regarded
as the temporal and spatial components of generalized four current
$\left\{ J^{^{µ}}\right\} =\left\{ \rho,\overrightarrow{J\,}\right\} $
of dyons. As such, we may write the generalized electromagnetic field
vector $\overrightarrow{\psi}$ in the following manner in terms of
components of generalized four potential $\left\{ V^{^{µ}}\right\} =\left\{ \varphi,\overrightarrow{V\,}\right\} $
with $\varphi=\phi_{e}-i\,\phi_{g}$ and $\overrightarrow{V\,}=\overrightarrow{A\,}-i\,\overrightarrow{B}$
as

\begin{equation}
\overrightarrow{\psi}=-\frac{\partial\overrightarrow{V}}{\partial t}-\overrightarrow{\nabla}\varphi-i\,\overrightarrow{\nabla}\times\overrightarrow{\psi}.\label{eq:16}\end{equation}
We may now write the following covariant forms of generalized Maxwell's-Dirac
equations of dyons i.e.

\begin{eqnarray}
\partial^{\nu}F_{\mu\nu} & =F_{\mu\nu,\nu} & =j_{µ},\,\,\,\,\,\,\,\,\,\partial^{\nu}\tilde{F}_{\mu\nu}=\tilde{F}_{\mu\nu,\nu}=k_{µ},\label{eq:17}\end{eqnarray}
with 

\begin{eqnarray}
F_{\mu\nu} & = & E_{\mu\nu}-\tilde{H}_{\mu\nu},\,\,\,\,\,\,\,\,\,\,\,\,\,\,\,\,\,\,\,\,\,\,\,\,\,\,\,\,\,\,\,\,\tilde{F}_{\mu\nu}=H_{\mu\nu}+\tilde{E}_{\mu\nu},\nonumber \\
E_{\mu\nu} & = & A_{\mu,\nu}-A_{\nu,\mu}=\partial_{\nu}A_{µ}-\partial_{µ}A_{\nu},\,\,\,\,\,\,\,\, H_{\mu\nu}=B_{\mu,\nu}-B_{\nu,\mu}=\partial_{\nu}B_{µ}-\partial_{µ}B_{\nu},\nonumber \\
\tilde{E}_{\mu\nu} & = & \frac{1}{2}\varepsilon_{\mu\nu\sigma\lambda}E^{^{\sigma\lambda}},\,\,\,\,\,\,\,\,\,\,\,\,\,\,\,\,\,\,\,\,\,\,\,\,\,\,\,\,\,\,\,\tilde{H}_{\mu\nu}=\frac{1}{2}\varepsilon_{\mu\nu\sigma\lambda}H^{^{\sigma\lambda}},\label{eq:18}\end{eqnarray}
where the tidle denotes the dual , $E_{\mu\nu}$ and $H_{\mu\nu}$
are the electromagnetic field tensors respectively due to the presence
of electric and magnetic charges and $\varepsilon_{\mu\nu\sigma\lambda}$
is the four index Levi - Civita symbol. Generalized electric and magnetic
fields of dyons given by equations (\ref{eq:13}) may now directly
be obtained from the components of field tensors $F_{\mu\nu}$ and
$\tilde{F}_{\mu\nu}$ described in terms of two potential as,

\begin{eqnarray}
F_{0j} & = & E^{^{j}},\,\,\,\,\,\,\,\,\,\,\,\,\,\,\,\,\,\,\, F_{jk}=\varepsilon_{jkl}H^{^{l}},\nonumber \\
\tilde{F}_{0j} & = & -H^{^{j}}\,\,\,\,\,\,\,\,\,\,\,\,\,\,\,\,\,\,\tilde{F}_{jk}=-\varepsilon_{jkl}E^{^{l}}.\label{eq:19}\end{eqnarray}
Taking the curl of second part of equation ( \ref{eq:15}) and using
first part of equation (\ref{eq:15}), we obtain a new vector parameter
$\overrightarrow{S}$ (say) as

\begin{eqnarray}
\overrightarrow{S} & = & \square\overrightarrow{\psi}=-\frac{\partial\overrightarrow{J}}{\partial t}-\overrightarrow{\nabla}\rho-i\,\overrightarrow{\nabla}\times\overrightarrow{J}\,.\label{eq:20}\end{eqnarray}
where $\square$ represents the D' Alembertian operator i.e.

\begin{equation}
\square=\frac{\partial^{2}}{\partial t^{2}}-\frac{\partial^{2}}{\partial x^{2}}-\frac{\partial^{2}}{\partial y^{2}}-\frac{\partial^{2}}{\partial z^{2}}\,.\label{eq:21}\end{equation}
Let us define the complex form of generalized field tensor of dyons
as 

\begin{equation}
G_{\mu\nu}=F_{\mu\nu}-i\,\tilde{F}_{\mu\nu},\label{eq:22}\end{equation}
which directly combines the two differential equations given by equation
(\ref{eq:17}) in to the following covariant field equation of dyons
i.e.

\begin{eqnarray}
\partial^{\nu}G_{\mu\nu} & =G_{\mu\nu,\nu} & =J_{µ},\label{eq:23}\end{eqnarray}
with $G_{\mu\nu}=V_{\mu,\nu}-V_{\nu,\mu}=\partial_{\nu}V_{µ}-\partial_{µ}V_{\nu}$
is named as the generalized electromagnetic field tensors of dyons.
The present model for generalized fields of dyons may also be explained
\cite{key-39,key-40} in terms of the suitable Lagrangian density
to obtain the field equation described above along with the following
form of Lorentz force equation of motion for dyons i.e.

\begin{eqnarray}
f_{µ}=m\,\frac{d^{2}x_{\mu}}{d\,\tau^{2}} & = & \frac{1}{2}\left(q\, G_{\mu\nu}^{\star}+q^{\star}G_{\mu\nu}\right)u^{\nu}=\left(e\, F_{\mu\nu}+g\,\tilde{F}_{\mu\nu}\right)u^{\nu}\label{eq:24}\end{eqnarray}
where $m$ is the mass of particle, $x_{\mu}$ is the position, $\tau$
is the proper time, $(\star)$ denotes the complex conjugate and $\{ u^{\nu}\}$
is the four-velocity of the particle. Equation (\ref{eq:24}) reduces
to the following usual form of equation of motion for a particle carrying
simultaneously electric and magnetic charges i.e.

\begin{equation}
m\,\overset{\overrightarrow{..}}{x}=\overrightarrow{f}=e\,(\overrightarrow{E}+\overrightarrow{u}\times\overrightarrow{H})+g\,(\overrightarrow{H}-\overrightarrow{u}\times\overrightarrow{E})\label{eq:25}\end{equation}
where $\overrightarrow{E}$ and $\overrightarrow{H}$ are electric
and magnetic fields given by equation (\ref{eq:13}). Equation (\ref{eq:25})
may also be written accordingly \cite{key-40} in terms of generalized
charge and generalized vector field $\overrightarrow{\psi}$.

\section{Electromagnetic Duality}

~~~~~~The concept of duality has been receiving much attention
\cite{key-17,key-18,key-19,key-20,key-49,key-50,key-51,key-52,key-53,key-54}
in gauge theories, field theories, supersymmetry and super strings.
Duality invariance is an old idea introduced a century ago in classical
electromagnetism for Maxwell's equations in vacuum as these were invariant
not only under Lorentz and conformal transformations but also invariant
under the following duality transformations,

\begin{eqnarray}
\overrightarrow{E} & = & \overrightarrow{E}\,\cos\theta+\overrightarrow{H}\,\sin\theta,\,\,\,\,\,\,\,\,\,\,\,\overrightarrow{H}=-\overrightarrow{E}\,\sin\theta+\overrightarrow{H}\,\cos\theta.\label{eq:26}\end{eqnarray}
Dirac \cite{key-1} put forward this idea and introduced the concept
of magnetic monopole not only to symmetrize Maxwell's equations but
also to make them dual invariant. Consequently, the the GDM equations
(\ref{eq:12}) are invariant under duality transformations (\ref{eq:26}).
For a particular value of $\theta=\frac{\pi}{2}$ we get the following
discrete duality transformations,

\begin{equation}
\overrightarrow{E}\,\,\mapsto\overrightarrow{H};\,\,\,\,\,\,\,\,\,\overrightarrow{H}\mapsto-\overrightarrow{E}.\label{eq:27}\end{equation}
and correspondingly we get the duality transformations for electric
and magnetic charges i.e.

\begin{equation}
e\,\mapsto g;\,\,\,\,\,\,\,\,\, g\mapsto-e.\label{eq:28}\end{equation}
Under these duality transformations the GDM equation (\ref{eq:12})
and equation (\ref{eq:25}) for the Lorentz force equation of motion
are invariant. As such, the duality is preserved in the theory of
simultaneous existence of electric and magnetic charges if we include
transformations (\ref{eq:27}) to the electric and magnetic fields
of dyons given by equation (\ref{eq:13}) along with the following
transformations to the potential and current components i.e.,\begin{eqnarray}
\left\{ A_{µ}\right\} \mapsto\left\{ B_{µ}\right\} ; & \left\{ B_{µ}\right\}  & \mapsto-\left\{ A_{µ}\right\} \nonumber \\
\left\{ j_{µ}\right\} \mapsto\left\{ k_{µ}\right\} ; & \left\{ k_{µ}\right\}  & \mapsto-\left\{ j_{µ}\right\} .\label{eq:29}\end{eqnarray}

In general we can write the duality transformations in terms of duality
matrix representation as

\[
\left(\begin{array}{c}
e\\
g\end{array}\right)=\left(\begin{array}{cc}
0 & -1\\
1 & 0\end{array}\right)\left(\begin{array}{c}
e\\
g\end{array}\right)\]

\begin{eqnarray}
\left(\begin{array}{c}
\overrightarrow{E}\,\\
\overrightarrow{H}\,\end{array}\right) & \mapsto\left(\begin{array}{cc}
0 & -1\\
1 & 0\end{array}\right) & \left(\begin{array}{c}
\overrightarrow{E}\,\\
\overrightarrow{H}\,\end{array}\right),\nonumber \\
\left(\begin{array}{c}
A_{µ}\\
B_{µ}\end{array}\right) & \mapsto\left(\begin{array}{cc}
0 & -1\\
1 & 0\end{array}\right) & \left(\begin{array}{c}
A_{µ}\\
B_{µ}\end{array}\right),\nonumber \\
\left(\begin{array}{c}
j_{µ}\\
k_{µ}\end{array}\right) & \mapsto\left(\begin{array}{cc}
0 & -1\\
1 & 0\end{array}\right) & \left(\begin{array}{c}
j_{µ}\\
k_{µ}\end{array}\right).\label{eq:30}\end{eqnarray}
Consequently, the covariant forms of GDM given by equations (\ref{eq:17})
are invariant under the transformations of duality along with the
following transformations for the components of electromagnetic field
tensors i.e.

\begin{eqnarray}
\left(\begin{array}{c}
F_{\mu\nu}\\
\tilde{F_{\mu\nu}}\end{array}\right) & \mapsto\left(\begin{array}{cc}
0 & -1\\
1 & 0\end{array}\right) & \left(\begin{array}{c}
F_{\mu\nu}\\
\tilde{F_{\mu\nu}}\end{array}\right)\label{eq:31}\end{eqnarray}
If we consider the generalized fields of dyons as complex quantities
the duality transformations now take the following forms respectively
for the electromagnetic complex field vector, generalized charge,
potential, current and generalized electromagnetic field tensors of
dyons as 

\begin{eqnarray}
\overrightarrow{\psi} & \,\,\,\,\,\mapsto & \exp i\theta(\overrightarrow{\psi})\nonumber \\
q & \mapsto & \exp i\theta\,(q)\nonumber \\
V_{µ} & \mapsto & \exp i\theta\,(V_{µ})\nonumber \\
J_{µ} & \mapsto & \exp i\theta\,(J_{µ})\nonumber \\
G_{\mu\nu} & \mapsto & \exp i\theta\,(G_{\mu\nu})\label{eq:32}\end{eqnarray}
Hence with these transformations equations the GDM equations (\ref{eq:15}),
covariant field equation (\ref{eq:23}) and equation of motion (\ref{eq:25})
associated with dyons in complex representation are invariant. The
energy and momentum densities of generalized electromagnetic field
of dyons is described respectively as

\begin{eqnarray}
\frac{1}{2} & \left|\overrightarrow{\psi}\right|^{^{2}} & =\frac{1}{2}(\left|\overrightarrow{E}\right|^{^{2}}+\left|\overrightarrow{H}\right|^{^{2}});\,\,\,\,\,\,\,\,\,\,\,\frac{1}{2i}(\overrightarrow{\psi})^{\star}\times(\overrightarrow{\psi)}=\overrightarrow{E}\times\overrightarrow{H}\label{eq:33}\end{eqnarray}
are invariant under duality transformations and the real and imaginary
parts of 

\begin{eqnarray}
\frac{1}{2}(\overrightarrow{\psi})^{^{2}} & = & \frac{1}{2}(\left|\overrightarrow{E}\right|^{^{2}}-\left|\overrightarrow{H}\right|^{^{2}})-i\,\overrightarrow{E}\cdot\overrightarrow{H}\label{eq:34}\end{eqnarray}
are described respectively as the Lagrangian and the charge density
density \cite{key-48} transform as the doublet under the duality
group. The duality is thus maintained only with the introduction of
magnetic charge which has been always a challenging new frontier and
its existence is still questionable.The present model is still consistent
for point like monopoles in view of covariant formulation and duality
invariance. The coupling between two generalized charges $q_{i}$
and $q_{k}$ is described as

\begin{equation}
q_{i}^{^{\star}}q_{k}=(e_{j}e_{k}+g_{j}g_{k})-i\,(e_{j}g_{k}-e_{k}g_{j})=\alpha_{jk}-i\,µ_{jk}\label{eq:35}\end{equation}
where the real part $\alpha_{jk}$ is called the electric coupling
parameter ( the Coulomb like term) responsible for the existence of
either electric charge or magnetic monopole while the imaginary part
$µ_{jk}$ is the magnetic coupling parameter and plays an important
role for the existence of magnetic charge. Both of these parameters
are invariant under the duality transformations. The parameter $µ_{jk}$
has also been named as Chirality quantization parameter \cite{key-2,key-3,key-39,key-40}
for dyons and leads the following charge quantization condition i.e.

\begin{eqnarray}
µ_{jk} & = & \pm n\,\,(n=0,1,2,3,.....)\label{eq:36}\end{eqnarray}
where the half integral quantization is forbidden by chiral invariance
and locality in commutator of the electric and magnetic vector potentials
\cite{key-39}. If we consider two dyons with $q_{j}=(e,0)$ and $q_{k}=(0,g)$
the quantization condition (\ref{eq:36}) reduces to well known Dirac
quantization condition $eg=\pm n\,$. If we do not consider Dirac
particle as dyon the Dirac quantization condition loses its dual invariance.
Thus dyon plays an important role in electromagnetic duality with
the association of Chirality quantization parameter and it is important
to consider the consistent quantum field theory for the simultaneous
existence of electric and magnetic charges (dyons). Now it is also
expected \cite{key-48} that just as the energy density given by equation
(\ref{eq:33}) respects the symmetry given by equation (\ref{eq:32})
for generalized vector field,on the similar grounds the mass formula
for dyon also respects this symmetry (\ref{eq:32}) for the generalized
charge so that

\begin{eqnarray}
M(e,g) & = & M(\left|e-i\, g\right|)=M\,\left(\sqrt{e^{^{2}}+g^{^{2}}}\right)\label{eq:37}\end{eqnarray}
which plays an important role in supersymmetric gauge theories.

\section{Quaternion Formulation for Dyons}

In section 3 we have reduced four sets of GDM (\ref{eq:12}) to two
sets of field equation (\ref{eq:13}) and we have established a relation
between potential and field by equation (\ref{eq:16}) and correspondingly
derived a relation between current and the new field vector given
by equation (\ref{eq:20}). These two equations are a kind of potential
and current equation and have a direct analogue to quaternion formulation
of dyons. To do this let us define the space time four differential
operator as the quaternion in the following manner 

\begin{eqnarray}
\boxdot & = & -i\,\frac{\partial}{\partial t}+e_{1}\frac{\partial}{\partial x}+e_{2}\frac{\partial}{\partial y}+e_{3}\frac{\partial}{\partial z}\label{eq:38}\end{eqnarray}

which has been the quaternionic form of four differential operator
$\left\{ \partial_{\mu}\right\} =\left(-i\,\frac{\partial}{\partial t}\,,\overrightarrow{\nabla}\right).$similarly
we can define quaternionic form of generalized four potential of dyons
as 

\begin{eqnarray}
V & = & -i\,\varphi+e_{1}V_{x}+e_{2}V_{y}+e_{3}V_{z}.\label{eq:39}\end{eqnarray}
Now operating equation (\ref{eq:38} ) to equation (\ref{eq:39})
and using the relations (\ref{eq:2}) for the quaternion units, we
get 

\begin{eqnarray}
\boxdot V & = & \psi\label{eq:40}\end{eqnarray}
where $\varphi$is again a quaternion defined as 

\begin{eqnarray}
\psi & = & -\psi_{t}+ie_{1}\psi_{x}+ie_{2}\psi_{y}+ie_{3}\psi_{z}\label{eq:41}\end{eqnarray}
with 

\begin{eqnarray}
\psi_{t} & = & \partial_{t}\varphi+\partial_{x}V_{x}+\partial_{y}V_{y}+\partial_{z}V_{z}.\label{eq:42}\end{eqnarray}
Here we have $\overrightarrow{\psi}=(\psi_{x},\psi_{y},\psi_{xz})$
and $\psi_{t}=o$ because of Lorentz gauge conditions applied separately
on electric and magnetic four potential. As such, the relation (\ref{eq:16})
represents the quaternion differential equation form of generalized
potential of dyons. Let us take the any inhomogeneous quaternion differential
equation as

\begin{eqnarray}
\overline{\boxdot}f & = & b\,\label{eq:43}\end{eqnarray}
where $\overline{\boxdot}=-i\,\frac{\partial}{\partial t}-e_{1}\frac{\partial}{\partial x}-e_{2}\frac{\partial}{\partial y}-e_{3}\frac{\partial}{\partial z}$
is the quaternion conjugate differential operator. This equation reduces
to the GDM equations (\ref{eq:15}) by identifying quaternion function
variable $f$ as the $\psi$ and the right hand side quaternion variable
as the quaternion valued four current of dyons given by

\begin{eqnarray}
J & = & -i\,\rho+e_{1}J_{x}+e_{2}J_{y}+e_{3}J_{z}.\label{eq:44}\end{eqnarray}
Similarly on operating equation (\ref{eq:38} ) to equation (\ref{eq:44})
we get 

\begin{eqnarray}
\boxdot J & = & S\label{eq:45}\end{eqnarray}
where 

\begin{eqnarray}
S & = & -S_{t}+ie_{1}S_{x}+ie_{2}S_{y}+ie_{3}S_{z},\nonumber \\
S_{t} & = & \partial_{t}\rho+\partial_{x}J_{x}+\partial_{y}J_{y}+\partial_{z}J_{z}.\label{eq:46}\end{eqnarray}
The components of $\overrightarrow{S}=(S_{x},S_{y},S_{z})$ are given
by equation (\ref{eq:20} ) and $S_{t}=\partial_{t}\rho+\partial_{x}J_{x}+\partial_{y}J_{y}+\partial_{z}J_{z}=0$
due to continuity equation applied individually on electric and magnetic
four currents. Thus we have seen that there are four sets of GDM differential
equation on real representation, two sets in complex representation
and one set in quaternion representation. Thus the quaternion formulation
of dyon is compact, simpler and manifestly covariant under quaternion
Lorentz transformation \cite{key-41} .The theory corresponds to two
dimensional representation in complex case and four dimensional representation
to real case. In other words $N=1$ quaternion representation maps
to $N=2$ dimensional complex and $N=4$ dimensional real representation.
Similarly, we may develop the quaternionic forms of other differential
equations in simple, compact and consistent manner. The theory of
quaternion variables presented here to the case of dyon is dual invariant
as the quaternion quantities are self dual. Here we have used the
bi quaternions instead of real quaternions to establish the relations
among the complex parameters of dyons. This model of quaternions is
valid for the classical and quantum dynamics of individual electric
and magnetic charges in the absence of each other. Thus two dual theories
are coupled together in this formalism and there is no difficulty
to represent them in terms of complex components of a bi quaternion
variables.

\section{Quaternion Dirac Equation for Dyons }

The free particle quaternion Dirac equation is described \cite{key-55,key-56}
as,

\begin{eqnarray}
(i\,\gamma^{\mu}\partial_{\mu}- & m & )\Psi(x,t)=0\label{eq:47}\end{eqnarray}
where $\Psi(x,t)=\left(\begin{array}{c}
\Psi_{a}(x,t)\\
\Psi_{b}(x,t)\end{array}\right)$ is the two component spinor and 

\begin{equation}
\Psi_{a}(x,t)=\Psi_{0}+i\,\Psi_{1};\,\,\,\,\,\,\,\,\,\,\,\Psi_{b}(x,t)=\Psi_{2}-i\,\Psi_{3}\label{eq:48}\end{equation}
are the components of spinor quaternion $\Psi$ and $\gamma$ matrices
are also defined in terms of quaternion units i.e.

\begin{eqnarray}
\gamma_{0} & = & \left[\begin{array}{cc}
1 & 0\\
0 & -1\end{array}\right];\,\,\,\,\,\,\,\,\,\,\,\,\gamma_{j}=e_{j}\left[\begin{array}{cc}
0 & 1\\
1 & 0\end{array}\right](j=1,2,3)\label{eq:49}\end{eqnarray}
The set of pure quaternion field defined by equation (\ref{eq:1}
) is invariant under the transformations 

\begin{eqnarray}
\phi & \rightarrow & \phi'=U\phi\overline{U},\,\,\,\,\,\,\,\,\,\, U\in Q,\,\,\,\, U\overline{U}=1\label{eq:50}\end{eqnarray}
and accordingly the quaternion conjugate transforms like 

\begin{eqnarray}
\overline{\phi'} & = & \overline{U\phi\overline{U}}=U\,\overline{\phi}\overline{U\,}=-U\phi\overline{U}=-\phi'.\label{eq:51}\end{eqnarray}
Any $U\in Q$ has a decomposition like equation (\ref{eq:50}) and
the quaternion differential equations of dyons discussed above are
invariant under these transformations of a quaternion. Transformation
equation (\ref{eq:50}) gives rise to a set $\{ U\in Q;\,\,\,\, U\overline{U}=1\}\sim SP(1)\sim SU(2).$
Though it has been emphasized earlier \cite{key-26,key-27,key-29}
that the automorphic transformation of $Q-$fields are local but one
can select it according to the representation. On the other hand a
$Q-$field is subjected to more general $SO(4)$ transformations 

\begin{eqnarray}
\phi & \rightarrow & \phi'=U_{1}\phi\overline{U}_{2},\,\,\,\,\,\,\,\,\,\, U_{1},U_{2}\in Q,\,\,\,\, U_{1}\overline{U_{1}}=U_{2}\overline{U_{2}}=1.\label{eq:52}\end{eqnarray}
and the covariant derivative is described in terms of two $Q-$ gauge
fields i.e

\begin{eqnarray}
D_{\mu}\phi & =\partial_{\mu}\phi & +A_{µ}\phi-\phi B_{µ}\label{eq:53}\end{eqnarray}
with

\begin{eqnarray}
A_{µ}' & = & U_{1}A_{µ}\overline{U_{1}}+(\partial_{\mu}U_{1})\overline{U_{1}};\nonumber \\
B_{µ}' & = & U_{2}B_{µ}\overline{U_{2}}+(\partial_{\mu}U_{2})\overline{U_{2}};\label{eq:54}\end{eqnarray}
where $A_{µ}$ and $B_{µ}$ are used in our model as the four potential
associates with electric and magnetic charge of dyons in Abelian gauge
theory where the gauge transformations are Abelian and global and
the quaternion covariant derivative given by equation ( \ref{eq:53}
) supports the idea of two four potential. As such it is possible
to extend our theory to the case of non-Abelian gauge theory and we
way write the Dirac equation for dyons as follows by replacing the
partial derivative to covariant derivative i.e. 

\begin{eqnarray}
(i\,\gamma^{\mu}D_{\mu}- & m & )\Psi(x,t)=0\label{eq:55}\end{eqnarray}
with the following definition

\begin{eqnarray}
\left[D_{\mu},D_{\nu}\right]\Psi & =D_{\mu}(D_{\nu} & \Psi)-D_{\nu}(D_{\mu}\Psi)=F_{\mu\nu}\Psi-\tilde{\Psi F_{\mu\nu}}\label{eq:56}\end{eqnarray}
where the gauge field strengths $F_{\mu\nu}$ and $\tilde{F_{\mu\nu}}$
are defined in section-3 for Abelian gauge theory of generalized electromagnetic
fields associated with dyons. Here we can express now the four potentials
(gauge potentials) in terms of quaternion as

\begin{eqnarray}
A_{µ} & = & A_{µ}^{^{0}}e_{0}+A_{µ}^{^{1}}e_{1}+A_{µ}^{^{2}}e_{2}+A_{µ}^{^{3}}e_{3},\nonumber \\
B_{µ} & = & B_{µ}^{^{0}}e_{0}+B_{µ}^{^{1}}e_{1}+B_{µ}^{^{2}}e_{2}+B_{µ}^{^{3}}e_{3}.\label{eq:57}\end{eqnarray}
As such, the Abelian theory of dyons discussed in section-3 and section-4
can now be restored by putting the conditions $U_{1}\overline{U_{1}}=U_{2}\overline{U_{2}}=1$
implying that $(A_{µ}^{^{0}})'=(A_{µ}^{^{0}})$ and $(B_{µ}^{^{0}})'=(B_{µ}^{^{0}})$.
It is also possible when we have $A_{µ}=\overline{A_{µ}}$ and $B_{µ}=\overline{B_{µ}}$.
However if we consider the imaginary quaternion i.e. $A_{µ}=-\overline{A_{µ}}$
and $B_{µ}=-\overline{B_{µ}}$ we have the $SU(2)\times SU(2)$ gauge
structure and $A_{µ}=A_{µ}^{^{a}}e_{a}=A_{µ}^{^{1}}e_{1}+A_{µ}^{^{2}}e_{2}+A_{µ}^{^{3}}e_{3}$
and $B_{µ}=B_{µ}^{^{a}}e_{a}=B_{µ}^{^{1}}e_{1}+B_{µ}^{^{2}}e_{2}+B_{µ}^{^{3}}e_{3}$.
Thus with the implementation of condition $U_{1}\overline{U_{1}}=U_{2}\overline{U_{2}}=1$
there are only the six gauge fields $A_{µ}^{^{a}}$and $B_{µ}^{^{a}}$
associated with the covariant derivative of Dirac equation (\ref{eq:55}).
The transformation equation (\ref{eq:52}) is continuous and isomorphic
to $SO(4)$i.e.

\begin{eqnarray}
\overline{\phi'}\phi' & = & \overline{(U_{1}\phi\overline{U}_{2})}(U_{1}\phi\overline{U}_{2})=U_{2}\overline{\phi}\,\overline{U_{1}}\, U_{1}\phi\overline{U_{2}}=U_{2}\overline{\phi}\phi\overline{U_{2}}=\overline{\phi}\phi.\label{eq:58}\end{eqnarray}
The resulting $Q-$ gauge theory has the correspondence $SO(4)\sim SO(3)\times SO(3)$
and the spinor transforms as left and right component (electric or
magnetic) spinors as 

\begin{eqnarray}
\Psi_{e} & \mapsto(\Psi_{e})'=U_{1} & \Psi_{e}\,\,\,\,\,\,\,\,,\nonumber \\
\Psi_{g} & \mapsto(\Psi_{g})'=U_{2} & \Psi_{g\,\,\,\,\,\,\,}.\label{eq:59}\end{eqnarray}
The following split basis of quaternion units my also be considered
as 

\begin{eqnarray}
u_{0} & = & \frac{1}{2}(1-i\, e_{3})\,\,\,,\,\,\,\,\,\, u_{0}^{\star}=\frac{1}{2}(1+i\, e_{3})\,\,\,,\nonumber \\
u_{1} & = & \frac{1}{2}(e_{1}+i\, e_{2})\,\,\,,\,\,\,\,\,\, u_{1}^{\star}=\frac{1}{2}(e_{1}-i\, e_{2})\,\,\,\label{eq:60}\end{eqnarray}
to constitute the $SU(2)$ doublets. As such, we may classify the
$Q-$classes into five groups and can describe the theory accordingly.
These five irreducible representations of $SO(4)$ are realized as 

\begin{eqnarray}
1. & (U_{1},U_{2}) & \Rightarrow SO(4)\mapsto(2,2)\nonumber \\
2. & (U_{1},U_{1}) & \Rightarrow SU(2)\mapsto(3,1)\nonumber \\
3. & (U_{2},U_{2}) & \Rightarrow SU(2)\mapsto(1,3)\nonumber \\
4. & (U_{1},1) & \Rightarrow Spinor\mapsto(2,1)\nonumber \\
5. & (U_{2},1) & \Rightarrow Spinor\mapsto(1,2)\label{eq:61}\end{eqnarray}
and accordingly it is easier to develop a non-Abelian gauge theory
of dyons. Here it is important to emphasize that the $\gamma-$ matrices
are defined above in terms of quaternions and when we talk about quaternion
local gauge transformations the $\gamma-$matrices are to be local
and space time dependent. Such possibility of $\gamma-$matrices is
already explored by Dehnen-Hitzer \cite{key-57} . Here we have just
described the existence of quaternion Dirac equation with out deriving
its solution and other consequences which will be further programme.

\section{Non-Abelian Dyons}

~~~~Now it is widely recognized that magnetic monopoles are better
understood in non-Abelian gauge theories but if we apply the two potential
approach to them, it leads to the existence of two different photons.
This possibility may be seen only in Abelian gauge theories not in
non Abelian gauge theories where it is not important to consider the
complex gauge potential. Here the gauge potentials are subjected by
one of the above mentioned five irreducible representations of quaternion
transformations. In Abelian gauge theory the existence of second photon
is minimized \cite{key-39,key-40} with the help of constancy condition
which relates the electric and magnetic parameters and is given by

\begin{eqnarray}
\frac{g}{e} & = & \frac{B_{µ}}{A_{µ}}=\frac{k_{µ}}{j_{µ}}=\frac{\tilde{F_{\mu\nu}}}{F_{\mu\nu}}=\tan\theta.\label{eq:62}\end{eqnarray}
Similar relation for $U_{1}$ and $U_{2}$ associated with quaternion
transformations may be developed. Equation (\ref{eq:62}) shows that
if dyons exists $\tan\theta$ should have finite value because for
$\theta=0$, the theory of electric charge particle exits while for
$\theta=\frac{\pi}{2}$ the theory of magnetic charge particle exists.
This may be one of the region that monopoles are not found as point
like objects but are considered as extended objects with finite shape
and size. First step in this direction for the theory of monopoles
was put forward by ' t Hooft-Polyakov \cite{key-8,key-9} who wrote
their famous paper on the existence of a regular magnetic monopole
solution in Georgi-Glashow model \cite{key-6} with the gauge group
$G=SO(3)$ is broken down to $U(1)$ by a Higgs field in the triplet
representation. Their argument implies the existence of such regular
monopoles in any unified gauge theory where the $U(1)$ of electromagnetism
would be a sub group of some simple non Abelian group. We can analyzed
this in terms of quaternion gauge groups too. The crucial difference
between non Abelian monopole and Dirac monopole was that these monopoles
appear as regular Solitons like solutions to the classical field equations
and can not be avoidable. This type of monopole satisfies the Dirac
quantization condition with the finite energy solutions. This may
be one of the important region with the definition of duality at $\theta=\frac{\pi}{2}$
where only monopole can exist with the different origin from the electric
charge. In nut shell we say that ' t Hooft-Polyakov monopole solutions
describe an object of finite size which from far away can not be distinguished
from a Dirac monopole of charge $g\mapsto-$$\frac{1}{e}$. If we
take the case $(U_{2},U_{2})\Rightarrow SU(2)\mapsto(1,3)$ ( isomorphic
to $SO(3)$ ) we can interpret our result analogous to the Georgi-Glashow
model consisting of $SO(3)$ gauge, $B_{µ}^{a}$ and Higgs triplet
$\Phi^{a}$ in the theory of ' t Hooft-Polyakov monopole and obtain
following equations of motion 

\begin{eqnarray}
D_{µ}F^{a\mu\nu}=\alpha\varepsilon^{abc}\Phi^{b}D^{\nu}\Phi^{c}\,\,, & (D_{\mu}D_{\mu}\Phi)^{a}= & -\lambda\Phi^{a}(\Phi^{b}\Phi^{b}-v^{2}),\nonumber \\
\,\, D_{\mu}\tilde{F}^{a\mu\nu} & = & 0\label{eq:63}\end{eqnarray}
where we introduce the Yang -Mills field strength 

\begin{eqnarray}
F_{\mu\nu}^{a} & = & \partial_{\mu}B_{\nu}^{a}-\partial_{\nu}B_{\mu}^{a}+\alpha\varepsilon^{abc}B_{\mu}^{b}B_{\nu}^{c},\label{eq:64}\end{eqnarray}
the covariant derivative 

\begin{eqnarray}
D_{µ}\Phi^{a} & = & \partial_{\mu}\Phi^{a}+\varepsilon^{abc}B_{\mu}^{b}\Phi^{c},\label{eq:65}\end{eqnarray}
and the Higgs potential 

\begin{eqnarray}
V(\Phi) & = & \frac{\lambda}{4}(\Phi^{a}\Phi^{a}-v^{2})^{2}.\label{eq:66}\end{eqnarray}
There are a simple formulas \cite{key-48} can be obtained accordingly
for the resultant masses of gauge particles

\begin{eqnarray}
M(e,0) & = & v\left|e\right|\label{eq:67}\end{eqnarray}
where $e$ is the eigen value of electric charge of a massive eigenstate
and $v$ specifies the the magnitude of the vacuum expectation value
of scalar Higgs field when we take $(U_{1},U_{1})\Rightarrow SU(2)\mapsto(3,1)$
structure of a quaternion. Similar formula can be calculated for '
t Hooft-Polyakov monopole named as Bogomolny bound \cite{key-58}
, is given by 

\begin{eqnarray}
M(0,g) & \geq & v\left|g\right|\label{eq:68}\end{eqnarray}
which is possible in Prasad-Sommerfield limit \cite{key-59} , where
the the ' t Hooft-Polyakov monopole and Julia - Jee dyon \cite{key-10}
solutions are generalized to vanish the self interaction of Higgs
field and in this case we obtain the Bogomolny equation $\mathbf{D\Phi=H}$
which is the reduction form of quaternion equation (\ref{eq:40})
. Olive \cite{key-48} showed that the Bogomolny bound (\ref{eq:68})
which is the self dual can be generalized to contain the higher value
of magnetic (topological) charge and thus it is possible for each
magnetic monopole Solitons to carry an electric charge $e$. The electromagnetic
duality hence gives rise the following bound on the mass of dyon for
all angles $\theta$ i.e.

\begin{eqnarray}
M & \geq & v\,(e\,\cos\theta+g\,\sin\theta).\label{eq:69}\end{eqnarray}
The sharpest bound is obtained when the right hand side has maximum
value and it happens for $e\,\sin\theta=g\,\cos\theta$. In other
words $\tan\theta=\frac{g}{e}$ which is the constancy condition (\ref{eq:62}).
Thus the electromagnetic duality given by equation (\ref{eq:32})
implies a generalization of Bogomolny bound (\ref{eq:68}) to give
rise the mass of a dyon as

\begin{eqnarray}
M(e\,,\, g) & = & v\,\left|e-i\, g\right|=v\,\sqrt{(e^{2}+g^{2}}),\label{eq:70}\end{eqnarray}
which is known as BPS mass formula. This mass formula does not distinguish
between the ´fundamental quantum particles and the magnetic monopoles,
being applicable to all of them, like meson-Solitons democracy in
Sine Gorden Model. The BPS mass formula is universal and is also invariant
under electromagnetic duality transformations.The natural frame work
for the realization of this symmetry in quantum field theory is the
$N=4$ supersymmetry \cite{key-60}.

\section{Duality in Supersymmetric Gauge Theories}

Finite quantum theories seem to exhibit a duality symmetry between
electricity and magnetism as stated earlier. The $\mathbb{Z}_{2}$duality
transforms the coupling constant of the theory and its inverse and
exchanges (non-Abelian) electric and magnetic charges. The spectrum
due to BPS mass formula given by equation (\ref{eq:70}) is invariant
under electromagnetic $\mathbb{Z}_{2}$ duality $(e,g)\mapsto(g,-e)$
is the consequence of the fact that the formula for Bogomolny bound
is invariant under duality and the spectrum saturates the bound. The
formula (\ref{eq:70} ) is actually invariant under the rotation in
the (e,g) plane but the quantization of the electric and magnetic
charges breaks the symmetry down to the $\mathbb{Z}_{2}$ duality
symmetry. This observation prompted Montonen and Olive \cite{key-49}
to conjecture that there should be a dual (´magnetic') description
of the gauge theory where the elementary particles are the BPS monopole
and the massive vector bosons appear as to be `electric monopole'.
It is speculated that the mass less Higgs could play the role of a
Goldstone boson associated to the breaking of $SO(2)$ symmetry down
to $\mathbb{Z}_{2}$. Montonen and Olive duality is expressed as\begin{eqnarray}
e\mapsto & g=\pm & \frac{1}{e}\,\,\,\,\,(in\,\, the\,\, units\,\, of\,\,4\pi\hbar).\label{eq:71}\end{eqnarray}
A dyon is not invariant under $CP-$transformation as we get

\begin{eqnarray}
(e,g) & \mapsto & (-e,g)\label{eq:72}\end{eqnarray}
and gauge theories contain a parameter that breaks $CP-$. This difficulty
is removed by Witten \cite{key-12} by adding a so called $\vartheta-$term
(or vacuum angle) to the Yang Mills Lagrangian with out spoiling its
renormalizability. Being a total derivative , it does not affect the
classical equations of motion. It violates $P-$and $CP-$but not
$C-$, which makes it as a good candidate for generalizing the long
range behaviour of the theory while maintaining the duality. By adding
the $CP-$violating term $\frac{e\vartheta}{8\pi^{2}}g$ to the electric
charge of dyon, the duality conjecture implies a richer dyonic spectrum
and has been tested accordingly. With this conjecture and applying
the Dirac quantization condition $eg=4\pi n_{m}$ where $n_{m}\in\mathbb{Z}$
, we get a general dyon as 

\begin{eqnarray}
(e,g) & \Rightarrow & (n_{e}e+\frac{e\vartheta}{2\pi}n_{m},\frac{4\pi}{e}n_{m})\,\,\,\,\,\,\,\,\,(\forall n_{e},n_{m}\in\mathbb{Z})\label{eq:73}\end{eqnarray}
where $n_{e},\&\,\, n_{m}$ are the numbers of electric and magnetic
charges presented in the system of a dyon respectively. Witten- effect
\cite{key-12} provides a physical meaning of electric charge $q_{e}=(n_{e}e+\frac{e\vartheta}{2\pi}n_{m})$
of a BPS monopole (dyon) to shift it by $\vartheta\mapsto\vartheta+2\pi$.
As such , the net charge on a dyon becomes

\begin{eqnarray}
Q & =(e,-i\, g)\mapsto & n_{e}e+\frac{e\vartheta}{2\pi}n_{m}-i\,\frac{4\pi}{e}n_{m}\Rightarrow\nonumber \\
Q=e\left[n_{e}+n_{m}(\frac{\vartheta}{2\pi}-i\,\frac{4\pi}{e^{2}})\right] & \Rightarrow & e(n_{e}+n_{m}\tau)\label{eq:74}\end{eqnarray}
where we have defined the complex parameter

\begin{eqnarray}
\tau & = & (\frac{\vartheta}{2\pi}-i\,\frac{4\pi}{e^{2}})\label{eq:75}\end{eqnarray}
and accordingly the Bogomolny bound takes the following form

\begin{eqnarray}
M & \geq & ve\left|n_{e}+n_{m}\tau\right|.\label{eq:76}\end{eqnarray}
To write the Lagrangian density in terms of complex parameter $\tau$
, it is now convenient to introduce the complex linear combination
of gauge field strengths in the similar manner as we have introduced
in the theory of dyons give by equation (\ref{eq:22} ) i.e.

\begin{eqnarray}
\overrightarrow{\mathfrak{\mathfrak{\boldsymbol{\mathrm{G_{\mu\nu}}}}}} & = & \mathbf{\overrightarrow{F_{\mu\nu}}-i}\overrightarrow{\tilde{\mathbf{F_{\mu\nu}}}}\label{eq:77}\end{eqnarray}
and we need accordingly the complex structure of gauge potential to
obtain this gauge field strength. This gauge field strength my be
obtained directly from quaternion gauge analyticity. When the 'theta
angle' is included in the complex coupling constant $\tau$ , the
group of duality transformation is $SL(2,\mathbb{Z})$ acting on the
vectors of electric and magnetic charges and projectively on the coupling
constant. 

It is to be noticed that the physics is periodic in $\vartheta$ with
a period $2\pi$ , we have the duality transformation ($\vartheta=2\pi$),

\begin{eqnarray}
T:\,\,\,\,\,\,\,\,\,\,\,\,\,\,\,\,\,\,\,\,\,\, & \tau\mapsto & \tau+1\label{eq:78}\end{eqnarray}
whereas the Montonen-Olive duality transformation (\ref{eq:71}) takes
the following form ( at $\vartheta=0$ ) in terms of $\tau$,

\begin{eqnarray}
S:\,\,\,\,\,\,\,\,\,\,\,\,\,\,\,\,\,\,\,\,\,\, & \tau & \mapsto-\frac{1}{\tau}.\label{eq:79}\end{eqnarray}
The operator $S$ and $T$ are the invertible operators and generate
a discrete group. Equations (\ref{eq:78} and \ref{eq:79} ) generate
the group $SL(2,\mathbb{Z})$ of projective transformations for full
duality transformations i.e.\begin{eqnarray}
\tau & = & \frac{a\tau+b}{c\tau+d}\,\,\,\,\,\,\,\,\,\,\, where\,\,\,\, a,b,c,d\in\mathbb{Z},\,\,\, and\,\,\, ad-bc=1\label{eq:80}\end{eqnarray}
BPS mass formula is thus invariant under $S$ and $T$ dualities and
plays an important role in the supersymmetric gauge theories as these
dualities are investigated for the invariance of supersymmetry and
super strings. We may now write 

\begin{eqnarray}
\left(\begin{array}{c}
n_{e}\\
n_{m}\end{array}\right) & \mapsto & \left(\begin{array}{cc}
a & -b\\
c & -d\end{array}\right)\left(\begin{array}{c}
n_{e}\\
n_{m}\end{array}\right),\label{eq:81}\end{eqnarray}
and the BPS bound now takes the form 

\begin{eqnarray}
M^{2} & \geq4v^{^{2}} & (n_{e},n_{m})\frac{1}{Im.\tau}\left[\begin{array}{cc}
1 & -Re.\tau\\
-Re.\tau & \left|\tau\right|^{2}\end{array}\right]\left(\begin{array}{c}
n_{e}\\
n_{m}\end{array}\right)\label{eq:82}\end{eqnarray}
which is invariant under $SL(2,\mathbb{Z})$ transformations. The
candidates for the invariance of $S$ and $T$ duality are the $N=2$
and $N=4$ extended supersymmetric gauge theories which could be established
respectively in terms of complex and quaternion representations.The
$N=2$ supersymmetry is crucial in that the electric and magnetic
charges enter the central charges in the supersymmetry algebra. It
is the external supersymmetry that explains the Bogomolny's bound
and provides an exact quantum status to BPS states. As we have stated
earlier that to write the GDM equations in compact simpler and self
dual representation we need to develop the theory of quaternion variables.
We can establish the generators of higher dimensional supersymmetry
\cite{key-61} by writing supercharge operators represented for $N=1$
, $N=2$ , $N=4$ supersymmetry respectively in terms of real number,
complex number and quaternion hyper complex number systems. Only these
numbers form a division ring and it is therefore impossible to extend
the supersymmetric gauge theories consistently beyond $N=4$ as the
the next number system octonions do not obey the law of associativity.
The manifold associated to $N=1,N=2,N=4$ gauge theories are also
related with the hyper complex number system namely Riemanian manifold
for $N=1$ (i.e. the algebra of real number), Kähler manifold for
$N=2$ (i.e. algebra of complex numbers) and the hyperkähler manifold
for $N=4$ supersymmetry describes from the algebra of quaternions.
This is the case when one describes one Chiral spinor representation
in $N=1$, two Majorana spinors for $N=2$ and four spinors (two Majorana
and two Weyl) in $N=4$ supersymmetric Yang Mills gauge theories.
Correspondingly we have one supersymmetric charge spinor operator
(only electric) in $N=1$ , two charge spinors in $N=2$ ,electric
and magnetic (dyon), but still incomplete to represent two component
spinors for both charges and a dyon with two spinors for both electric
and magnetic charges in $N=4$ supersymmetric gauge theories to represent
the electromagnetic as well as $S$ and $T$ duality of $SL(2,\mathbb{Z})$
gauge group. It is also believed that $N=4$ supersymmetric gauge
theory is the complete structural theory to explain the dyon mass
formula and supersymmetric generators represented therein  have the
direct one to one relation with this mass formula. It has already
been shown \cite{key-61} that the $\alpha$ and $\beta$ matrices
introduced by Osborn \cite{key-62} for the description of $N=4$
supersymmetric gauge theory have the same commutation and anti-commutation
relations as that of quaternion elements. A quaternion may also be
represented in terms of two complex structures of duality together
in quantum field and supersymmetric gauge theories. So, we have tried
to reformulate the duality and supersymmetric gauge theories in terms
of hyper complex numbers over the fields of real, complex and quaternion
number system. As such, it may be concluded that the quaternion formulation
be adopted in a better way to understand the explanation of the duality
conjecture and supersymmetric gauge theories as the candidate for
the existence of monopoles and dyons where the complex parameters
be described as the constituents of quaternion.

\textbf{Acknowledgment}: - We are thankful to German Academic Exchange
Service (DAAD), Bonn, Germany for providing financial support to O.
P. S. Negi to carry out this work at the Universität Konstanz under
re-invitation programme.

\end{document}